\documentclass[a4paper,aps,,floatfix]{revtex4}
\usepackage[dvips]{graphicx}
\usepackage{bm,isolatin1}
\begin{document}

\title{Dynamin recruitment by clathrin coats: a physical step?\\---\\
Recrutement des dynamines par les puits recouverts de clathrines : une
\'etape physique\,?}

\author{J.-B. Fournier}
\altaffiliation{Author for correspondence (jbf@turner.pct.espci.fr)}
\author{P.-G. Dommersnes}
\altaffiliation{Present address: Institut Curie, UMR 168, 26 rue d'Ulm,
F-75248 Paris Cédex 05, France.}
\affiliation{Laboratoire de Physico-Chimie Th\'eorique and FR CNRS 2438
``Mati\`ere et Syst\`emes Complexes'', ESPCI, 10 rue Vauquelin, F-75231
Paris cedex 05, France}
\author{P. Galatola}
\affiliation{LBHP, Universit\'e Paris 7---Denis Diderot and FR CNRS 2438
``Mati\`ere et Syst\`emes Complexes'', Case 7056, 2 place Jussieu,
F-75251 Paris cedex 05, France}

\date{\today}

\begin{abstract}
\smallskip
\textbf{Abstract} -- Recent structural findings have shown that dynamin,
a cytosol protein playing a key-role in clathrin-mediated endocytosis,
inserts partly within the lipid bilayer and tends to self-assemble
around lipid tubules. Taking into account these observations, we make
the hypothesis that individual membrane inserted dynamins imprint a
local cylindrical curvature to the membrane. This imprint may give rise
to long-range mechanical forces mediated by the elasticity of the
membrane.  Calculating the resulting many-body interaction between a
collection of inserted dynamins and a membrane bud, we find a regime in
which the dynamins are elastically recruited by the bud to form a collar
around its neck, which is reminiscent of the actual process preempting
vesicle scission. This physical mechanism might therefore be implied in
the recruitment of dynamins by clathrin coats.  

\medskip 
\noindent\textbf{endocytosis / clathrin / dynamin / membrane inclusions
interactions}

\bigskip
\textbf{Résumé} -- Des donnés structurales récentes ont montré que la
dynamine, une protéine du cytosol qui joue un rôle clé dans l'endocytose
clathrine-dépendante, s'insère partiellement dans la bicouche
membranaire et tend à s'auto-assembler autour de tubules lipidiques. En
tenant compte de ces observations, nous faisons l'hypothèse que les
dynamines impriment localement une courbure cylindrique dans la
membrane. Cette empreinte peut engendrer des forces élastiques de longue
portée. En calculant l'interaction multi-corps entre un ensemble de
dynamines insérées dans la membrane et une capsule endocytotique, nous
trouvons un régime dans lequel les dynamines sont recrutées
élastiquement par la capsule pour former un collier autour de son cou,
ce qui rappelle le processus précédant la scission des vésicules
d'endocytose. Ce mécanisme physique pourrait donc être impliqué dans le
recrutement des dynamines par les capsules de clathrine.

\medskip
\noindent\textbf{endocytose / clathrine / dynamine / interactions entre
inclusions membranaires}

\end{abstract}

\maketitle


\section{Introduction}\label{intro}

In eukaryotic cells, membranes of different organelles are functionally
connected to each other via vesicular transport. Formation of transport
vesicles from invaginated buds of the plasma membrane is called
\textit{endocytosis}~\cite{lodish_book}. In clathrin-mediated
vesiculation, vesicle formation starts with the assembly on the donor
membrane of a highly organized ``coat'' of clathrins~\cite{smith98},
which acts both to shape the membrane into a bud and to select cargo
proteins~\cite{pearse90,rothman94,schekman96,marsh99,takei01}. The
mechanism by which an invaginated clathrin-coated bud is converted to a
vesicle (scission) involves the action of a cytoplasmic GTPase protein
called \textit{dynamin}~\cite{schmid98,marsh99}. Dynamins form
oligomeric rings at the neck of deeply invaginated membrane buds and
induce scission~\cite{hinshaw95,takei95}. How exactly dynamin is
recruited and how the scission actually occurs remains
unclear~\cite{hinshaw00,danino01}. In this paper we propose that dynamin
recruitment by clathrin coats could be driven by long-ranged
\textit{physical} forces mediated by the membrane curvature elasticity.

Cryo-electron microscopy has recently revealed the detailed structure of
the clathrin coats at $21\,\mathrm{\AA}$ resolution~\cite{smith98}.
Clathrin units, also called ``triskelions'', have a star-like structure
with three legs. Initially solubilized into the cytoplasmic fluid, they
self-assemble onto the membrane surface into a curved, two-dimensional
solid scaffold. The latter is a honeycomb made of hexagons and pentagons
(geometrically providing the curvature) the sides of which are built by
the overlapping legs of the clathrin triskelions. In the plasma
membrane, clathrins usually interact with ``adaptor'' transmembrane
proteins, which also serve to select cargo proteins.  However it has
been shown that clathrin coats can readily self-assemble onto
protein-free liposomes~\cite{takei98,huang99}.

Dynamin is known to be solubilized in the cytosol as tetramers, and to
aggregate in low-salt buffers into rings and spirals~\cite{hinshaw95}.
Dynamin also self-assembles onto lipid bilayers, forming helically
striated tubules that resemble the necks of invaginated buds (tube
diameter $\simeq\!50\,\mathrm{nm}$)~\cite{takei95}.  Addition of GTP
induces morphological changes: either the tubules constrict and
break~\cite{sweitzer98}, or the dynamin spiral
elongates~\cite{stowell99}. These findings suggest that the
scission of clathrin-coated buds is produced by a mechanochemical
action~\cite{stowell99,marks01}. 

\begin{figure}
\centerline{\includegraphics[width=.4\columnwidth]{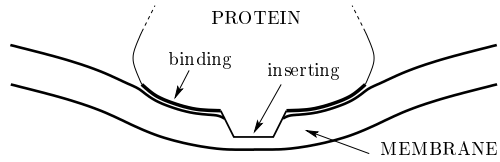}}
\caption{Schematic representation of a cytosol protein partly inserting within a
lipid bilayer and inducing a local membrane curvature via a binding
region.}
\label{incl}
\end{figure}

At the earlier stages of the budding process, dynamins already strongly
interact with bilayer membranes. Indeed, \textit{in
vivo} studies showed that dynamin binds acidic phospholipids in a way
that is essential to its ability to form oligomeric rings on invaginated
buds~\cite{tuma93,liu94,lin97,vallis99,lee99}.  Using a model lipid
monolayer spread at the air-water interface, it was shown that dynamins
actually penetrate within the acyl region of the membrane
lipids~\cite{burger00}. This finding was recently confirmed by the
three-dimensional reconstruction of the dynamin structure by
cryo-electron microscopy at $20\,\mathrm{\AA}$
resolution~\cite{zhang01}: dynamins form T-shaped dimers the ``leg'' of
which inserts partly into the outer lipid leaflet.

It was long ago suggested~\cite{gruler75,gruler77} that particles
inserted within bilayers should feel long-range interactions mediated by
the elasticity of the membrane. Indeed, a protein penetrating within a
bilayer and binding its lipids---such as dynamin---may in general
produce a local membrane curvature (see Fig.~\ref{incl}).  Because of
the very nature of the curvature elasticity of fluid membranes, this
deformation relaxes quite slowly away from its source, and the presence
of another inserted particle produces an interference implying an
interaction energy~\cite{gruler75}.  This holds as long as the
separation between the inclusions is smaller than the characteristic
length $\xi_\sigma=\sqrt{\kappa/\sigma}$, where
$\kappa\simeq60\,k_\mathrm{B}T$ is the bending rigidity of the membrane
and $\sigma$ is the tension of the membrane. At separations larger than
$\xi_\sigma$, the membranes flattens out and the interaction vanishes
exponentially. Note that the membrane tension is not a material constant
like $\kappa$; it is an effective force per unit area, which is most
probably biologically regulated~\cite{morris01} and is usually of the
order of $10^{-2}$ to $10^{-5}$ times the surface tension of ordinary
liquids~\cite{morris01,safran02}. At this point, anticipating on our
model for the clathrin-dynamin system, let us state
that we shall formally assume $\sigma=0$ in this paper, which amounts
to assuming that the relevant distances between the inclusions (i.e.,
the distance between the neck of the clathrin bud and the dynamins) are
smaller than $\xi_\sigma$. This means that our model should rather apply
to weakly tense membranes, e.g.,
$\sigma\simeq10^{-3}\,\mathrm{mJ}/\mathrm{m}^2$, for which
$\xi_\sigma\simeq400\,\mathrm{nm}$ (which is quite larger than the
typical size of the clathrin buds $\simeq80\,\mathrm{nm}$). Although the
actual value of $\sigma$ in the vicinity of clathrin buds is unknown,
such a small tension agrees with recent measurements on biological
membranes (erythrocytes membranes interacting with their
cytoskeleton)~\cite{safran02}. In the case of stronger tensions, we
expect our results to hold nonetheless, the dynamins being ``captured''
when their Brownian diffusion brings them at a distance from the bud
less than $\xi_\sigma$.

The first detailed calculation of the membrane mediated interaction was
performed for two isotropic particles each locally inducing a
\textit{spherical} curvature~\cite{goulian93,fournier97}.
The interaction was found to be repulsive, proportional to the
rigidity~$\kappa$ of the membrane and to the sum of the squares of the
imposed curvatures; it decays as $R^{-4}$, where $R$ is the distance
between the particles. The case of anisotropic particles producing
non-spherical membrane deformations is even more interesting, since
their collective action on the membrane is expected to have nontrivial
morphological consequences~\cite{bouligand90,fournier96,kralj-iglic99}.
The local deformation of a membrane actually involves \textit{two}
distinct curvatures, associated with two orthogonal directions (as in a
saddle or in a cylinder). Recent calculations showed that the
interaction between two anisotropic inclusions is very long-ranged and
decays as $R^{-2}$~\cite{park96,dommersnes99,chou01}. It is always
attractive at large separations and favors the orientation of the axis
of minor curvature along the line joining the
particles~\cite{dommersnes99}. Note that these elastic interactions
prevail at large separations, since they are of much longer range than
other forces, such as van der Waals or screened electrostatic
interactions. 

\section{Model}\label{model}

Among the above informations, let us outline the three points that are
essential for our model. (i) Clathrin coats are solid scaffolds that
rigidly shape extended parts of the membrane into spherical caps. (ii)
Dynamins are solubilized proteins that partly insert within the membrane
bilayer. (iii) Inserted membrane hosts that imprint a local membrane
curvature interact with long-range forces of elastic origin. 

\begin{figure*}
\centerline{
\includegraphics[width=.75\columnwidth]{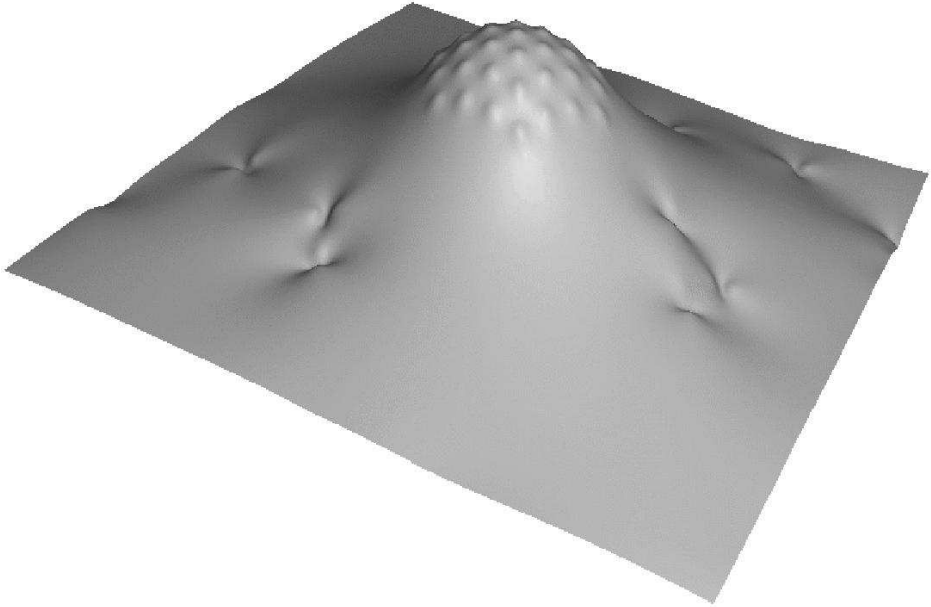}}
\caption{Piece of a model membrane showing a clathrin coated bud and the
imprints of inserted dynamins.}
\label{image_model}
\end{figure*}

We therefore build the following model. We consider a clathrin coated
bud as being a membrane patch bearing a constant, fixed spherical
curvature.  Technically, we shall build the bud by placing a large
number of point-like spherical curvature sources at the vertex of a
hexagonal lattice (see Fig.~\ref{image_model}). Within the present
formalism, this is the simplest way to define a rigid, almost
non-deformable, spherically curved zone. Since the dynamins will not
penetrate the bud, we expect that our results will not depend on whether
the bud is geometrically enforced (which would be conceptually simpler
but technically harder here) or built by an inclusion array.  Note that the
shape of the neck around the clathrin bud will result from the
minimization of the total energy, and will therefore not be enforced
artificially. 

Because dynamins partly insert within the membrane and seem to
accommodate cylindrical curvature, we model them as sources locally
imprinting a cylindrical curvature. We place a large number of such
``dynamins'' on a membrane in the presence of an artificial bud as
described above (Fig.~\ref{image_model}), and we study whether the
latter will recruit or not the dynamins through elastic long-range
forces.

\subsection{Long-range elastic interactions between many membrane
inclusions}\label{multi}

The elastic interaction between $N$ isotropic or anisotropic membrane
hosts can be calculated from first
principles~\cite{dommersnes99,kim99,kim00,dommersnes02}. The membrane is
described as a surface which is weakly deformed with respect to a
reference plane.  Without this assumption, analytical calculations are
virtually impossible. An obvious consequence is that we can accurately
describe only weakly invaginated buds; nevertheless, we expect that our
results will hold qualitatively for strongly invaginated buds. The
membrane hosts are described as point-like sources bearing two
curvatures $c_1$ and $c_2$, associated with two orthogonal directions.
These values represent the two principal curvatures that the hosts
imprint on the membrane through their binding with the membrane lipids
(assuming the binding region is itself curved). For instance, a
spherical impression corresponds to $c_1/c_2=1$, a cylindrical
impression corresponds to $c_1/c_2=0$, and a saddle-like impression
corresponds to $c_1/c_2=-1$.  This over-simplified model actually
contains the essential ingredients responsible for the long-range
elastic interactions between membrane inclusions: for protein hosts of a
size comparable to the thickness of the membrane, it yields accurate
interactions for separations as small as about three times the particles
size. Note that the curvature actually impressed by a particle could be
affected by the vicinity of another inclusion, we shall however neglect
this effect for the sake of simplicity (strong binding hypothesis). 

The point-like curvature sources describing the membrane hosts diffuse
and rotate within the fluid membrane, because of the forces and torques
exerted by the other membrane hosts and of the thermal agitation
$k_\mathrm{B}T$. We parameterize the orientation of a particle by the
angle $\theta$ that its axis of minor curvature, i.e., the axis
associated with min$(|c_1|,|c_2|)$, makes with the $x$-axis, in
projection on the $(x,y)$ reference plane.  Given $N$ inclusions with
specified positions $x_i$ and~$y_i$, orientations~$\theta_i$, and
curvatures $c_{1i}$ and~$c_{2i}$, for $i=1\ldots N$, we calculate the
shape of the membrane satisfying the $N$ imposed point-like curvatures
and we determine the total elastic energy of the system. We thereby
deduce the $N$-body interaction between the hosts
$F_\mathrm{int}(\ldots, x_i, y_i, \theta_i, c_{1i}, c_{2i}, \ldots)$.
The mathematical details of this procedure are sketched in
Appendix~\ref{app_calc}.

\subsection{Pairwise interactions}

\begin{figure}
\includegraphics[width=.5\columnwidth]{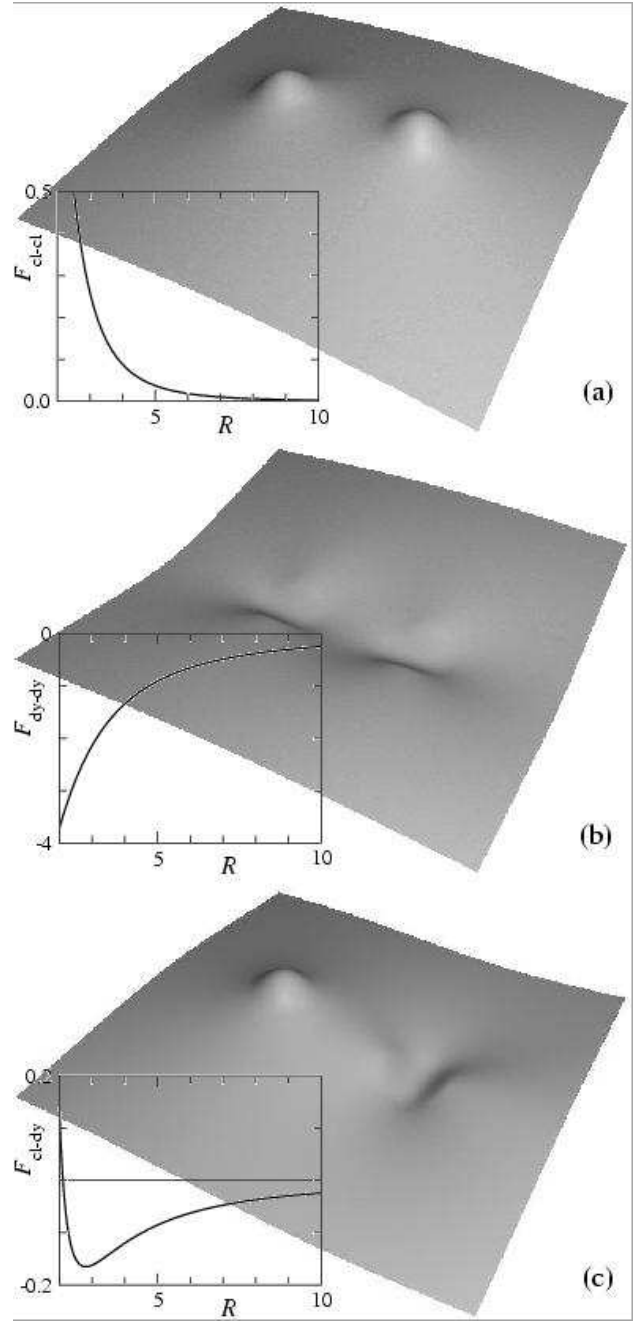}
\caption{Shape of a membrane distorted by two inclusions imprinting
local curvatures and interaction energy as a function of separation.
Distances are rescaled by the membrane thickness~$a$ and energies by
$\kappa a^2c^2$. (a) spherical inclusions of curvature $c$; (b)
cylindrical inclusions of curvature $c$; (c) spherical inclusion of
curvature $c$ and cylindrical inclusion of curvature $0.2\,c$. The shapes
are calculated from Eq.~(\protect\ref{ueq}), the interaction energies
from Eq.~(\protect\ref{interaction}).}
\label{fig:pairwise}
\end{figure}

Before studying the collective interaction between model dynamins and
clathrin coats, let us describe how point-like spherical and cylindrical
sources interact pairwise (in the absence of membrane tension, as
discussed in Sec.~\ref{intro}).

The membrane distortion produced by two inclusions modeled as point-like
spherical curvature sources is shown in Fig.~\ref{fig:pairwise}a.  Each
inclusion appears as a small spherical cap away from which the membrane
relaxes to a flat shape. As evidenced by the plot of the interaction
energy (see Fig.~\ref{fig:pairwise}a), such spherical inclusions repel
one another. Calling $\kappa$ the bending rigidity of the membrane, $a$
the thickness of the membrane (which is comparable to the size of the
inclusions), $c$ the curvature set by the inclusions and $R$ their
separation, our calculation gives (see Appendix~\ref{app_calc}):
\begin{equation}\label{Fclcl}
F_\mathrm{cl-cl}(R)\simeq8\pi\kappa\,a^2c^2\left(\frac{a}{R}\right)^4
\end{equation}
for the the asymptotic interaction at large separations, in agreement
with previous works~\cite{goulian93,park96,dommersnes99}. Note that the
plot given in Fig.~\ref{fig:pairwise}a corresponds to the exact interaction
within our model, not to the asymptotic expression~(\ref{Fclcl}).

The membrane distortion produced by two inclusions modeled as
point-like cylindrical curvature sources is shown in
Fig.~\ref{fig:pairwise}b. Each inclusion appears as a small cylindrical cap
away from which the membrane relaxes to a flat shape. The interaction
between two such hosts depends on their relative orientation. The
minimum energy is found when the axes of the cylinders are parallel to the
line joining the inclusions. As evidenced by the plot of the interaction
energy (see Fig.~\ref{fig:pairwise}b), the interaction is then
attractive. It therefore turns out that two such hosts produce a weaker
membrane deformation when they are close to one another than when they
are far apart. As described in Ref.~\cite{dommersnes99}, when their
curvature is strong enough, such inclusions tend to aggregate and to
form linear oligomers. Their asymptotic interaction energy is given by
\begin{equation}\label{Fdydy}
F_\mathrm{dy-dy}(R)\simeq-8\pi\kappa\,a^2c^2\left(\frac{a}{R}\right)^2.
\end{equation}
It decays as $R^{-2}$, hence it is of longer range than~(\ref{Fclcl}).

Finally, we show in Fig.~\ref{fig:pairwise}c the membrane distortion
produced by the interaction between a spherical source and a cylindrical
one. The latter is oriented in the direction that minimizes the
energy. As evidenced by the plot of Fig.~\ref{fig:pairwise}c, the interaction
is attractive at large separations and repulsive at short separations,
with a stable minimum configuration at a finite distance. Calling $c$
the curvature set by the cylindrical inclusion and $c'$ the one set
by the spherical inclusion, our calculations give the asymptotic
interaction
\begin{equation}\label{Fdycl}
F_\mathrm{cl-dy}(R)\simeq-4\pi\kappa\,a^2c\,c'
\left(\frac{a}{R}\right)^2.
\end{equation}
We may therefore expect that dynamins will be attracted by clathrins
coats; however, owing to the non-pairwise character of the
interaction~\cite{dommersnes99}, it is necessary to actually perform the
corresponding many-body calculation. It is also necessary to check
whether thermal agitation will or will not disorder the inclusions.

\section{Collective interactions between model dynamins and clathrin
buds}\label{collective}

As described in Sec.~\ref{model}, we build a model clathrin-coated bud
by placing in a membrane $N_\mathrm{cl}$ point-like spherical inclusions
of curvature $c_\mathrm{cl}$ on a hexagonal array with lattice
constant~$b$.  Here, we have chosen $N_\mathrm{cl}=37$ and $b=3a$. By
changing the curvature $c_\mathrm{cl}$, we can adjust the overall
curvature of the clathrin scaffold, thereby simulating the growth of a
vesicular bud. We then add $N_\mathrm{dy}=40$ point-like cylindrical
sources of curvature $c_\mathrm{dy}$ modeling inserted dynamins.

\begin{figure}
\centerline{\includegraphics[width=.44\columnwidth]{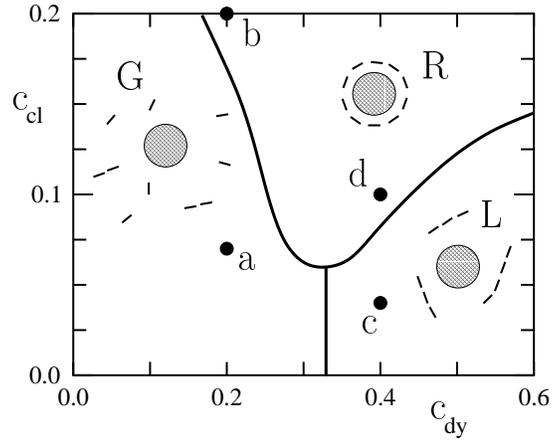}}
\caption{Phase diagram representing the typical equilibrium
configurations of a system of model dynamins in the vicinity of a curved
clathrin scaffold.  $c_\mathrm{dy}$ and $c_\mathrm{cl}$ are the
curvatures associated with the dynamins and clathrins, respectively, in
units of the inverse membrane thickness~$a$. In region $G$, the dynamins
form a ``gas'' non interacting with the clathrin scaffold. In region
$L$, the dynamins form a system of linear oligomers non interacting with
the clathrin scaffold. In region $R$, the dynamins form a ring around
the clathrin scaffold, which is reminiscent of real endocytosis.}
\label{phase_diag} 
\end{figure}

To study the collective behavior of this system under the action of the
multibody elastic interactions (see Sec.~\ref{multi}) and of thermal
agitation, we perform a Monte Carlo simulation. The details of the
simulations are given in Appendix~\ref{app_mcarl}. To prevent unphysical
divergences of the elastic interaction energy, it is necessary to
introduce a hard-core steric repulsion preventing two inclusions to
approach closer than a distance~$d$. Since the size of the inclusions
imprints is of the order of the membrane thickness~$a$, we have chosen
$d=2a$. Actually, at such microscopic separations, other short-ranged
interactions intervene, the details of which are still unknown. Here, we
disregard them, since our interest lies in the mechanism by which the
recruitment process and the formation of dynamin collars is driven. In a
later stage, which we do not model here, dynamin rings are further
stabilized by bio-chemical interactions~\cite{hinshaw00}.

The results of the Monte Carlo simulations are summarized in the phase
diagram of Fig.~\ref{phase_diag}, in terms of the curvatures
$c_\mathrm{cl}$ and $c_\mathrm{dy}$ of the clathrins and dynamins
imprints, respectively. Here, we have chosen to span $c_\mathrm{cl}$
between $0$ to $0.2\,a^{-1}$: for a lattice constant $b=3a$ and assuming
$a\simeq40$\,\AA, this corresponds for the clathrin-coated bud to a
maximum curvature of radius $\rho\simeq
b/(a\,c_\mathrm{cl})\simeq60\,\mathrm{nm}$. Since our clathrin patch has $7$
spherical sources on its diameter, the size of the bud is
$7b\simeq85\,\mathrm{nm}$. These values are typical for
clathrin-mediated endocytosis~\cite{marsh99}. For the dynamins, we have
spanned $c_\mathrm{dy}$ between $0.1\,a^{-1}$ and $0.4\,a^{-1}$, which
corresponds to a maximum curvature of the
imprint~$\simeq0.1\,\mathrm{nm}^{-1}$. As for the membrane bending
rigidity, we have taken $\kappa=60\,k_\mathrm{B}T$, since for biological
membranes at room temperature $\kappa$  lies between $50$ and $100$
$k_\mathrm{B}T$~\cite{song93,strey95}.

\begin{figure}
\includegraphics[width=.375\columnwidth]{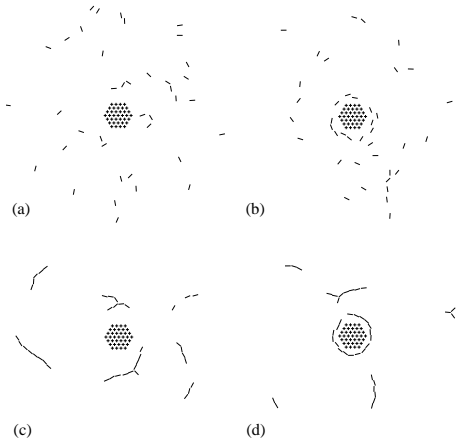}
\caption{Typical snapshots showing the equilibrium arrangement of model
dynamins (bars) in the vicinity of the clathrin scaffold (hexagonal
array). The figures (a), (b), (c) and (d) refer to the corresponding
points in the phase diagram of Fig.~\protect\ref{phase_diag}.}
\label{snapshots}
\end{figure}

\begin{figure*}
\centerline{
\includegraphics[width=.75\columnwidth]{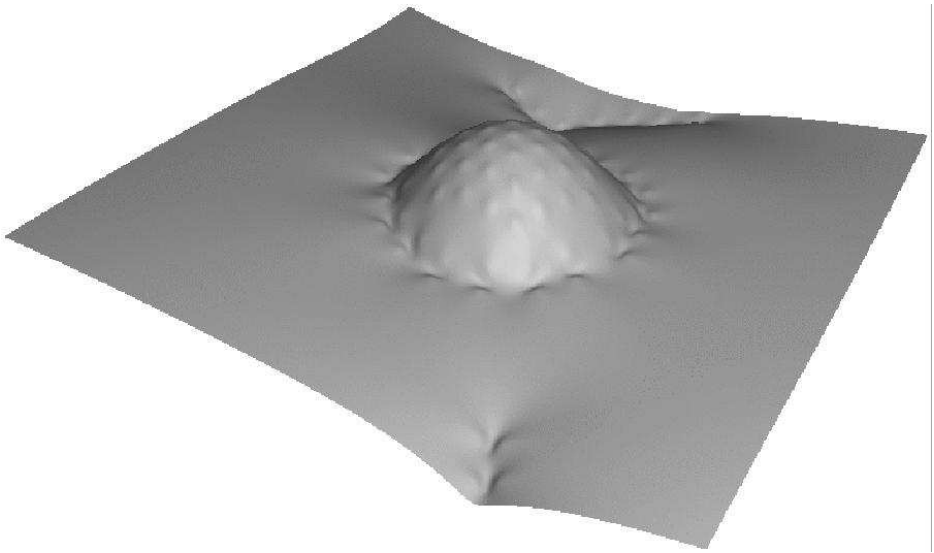}}
\caption{Membrane shape corresponding to point (d) in Figs.\
\protect\ref{phase_diag} and~\protect\ref{snapshots}, showing the
self-assembly of a ring of dynamins around a clathrin bud.}
\label{recrutement}
\end{figure*}

The phase diagram displays three regimes (see Fig.~\ref{phase_diag}): a
state in which the dynamins are disordered in a gas-like fashion ($G$),
a state in which the dynamins form linear oligomers that do not interact
with the clathrin bud ($L$), and a state in which the dynamins form a
ring around the clathrin bud ($R$). In region ($R$), due to the
shallowness of the dynamin imprints, the system is disordered by thermal
agitation. Increasing the curvature of the dynamin imprints increases
the elastic attraction between the dynamins (see
Fig~\ref{fig:pairwise}b) and leads to the formation of linear oligomers
($L$). These oligomers wrap around the clathrin bud when the latter
is sufficiently developed ($R$). Typical snapshots corresponding to the
four points $a$, $b$, $c$, $d$ in Fig.~\ref{phase_diag} are shown in
Fig.~\ref{snapshots}. Note that in Fig.~\ref{snapshots}b the dynamin
collar is rather ``gaseous'' due to the weakness of the dynamins'
imprints, while in Fig.~\ref{snapshots}d the ring is tight and
well-ordered.

It is difficult within the present paper to discuss the nature of the
boundaries between the different ``phases'' displayed in
Fig.~\ref{phase_diag}. Because we are considering a rather small
(thermodynamically speaking) number of dynamins, and because they are
under the influence of a localized curvature field, the transition lines
between the various regions of Fig.~\ref{phase_diag} are actually broad.
Attempting to describe them as first- or second-order
transition lines should not be meaningfull.

\section{Conclusion}

In this paper, we have shown that if membrane inserted dynamins produce
cylindrical imprints and if the latter are sufficiently curved, then the
resulting long-range forces mediated by the membrane elasticity are
strong enough to overcome Brownian motion and bring them into a collar
around the neck of a clathrin bud. Of course, simple diffusion could
also bring dynamins around clathrin buds, and their binding into a
ring could be the result of specific bio-chemical interactions. However,
if a cylindrical imprint can speed up this process, then evolution may
have selected it.

To test this model, one might look experimentally for linear oligomers
of dynamins (see Fig.~\ref{snapshots}c). However, since ``gaseous''
rings are also possible (see Fig.~\ref{snapshots}b), the existence of
such linear aggregates may not be necessary. It would be more
interesting to directly check, e.g., by cryo-electron microscopy, the
shape of the dynamin region that penetrates within the membrane. 

Finally, note that our model is obviously over-simplified: i)~many
other integral proteins float around dynamins, ii)~dynamins may interact
with various lipidic domains within the bilayer, iii)~the membrane may
have a spontaneous curvature due to its asymmetry, iv)~fluctuations are
not only thermal but also active, and hence could be larger than we
estimate, and v)~large values of the membrane tension could shorten the
range at which the dynamins are recruited (see Sec.~\ref{intro}).
Nonetheless, we believe that our model correctly captures the effects of
the anisotropic elastic interactions.

\begin{acknowledgments}
We acknowledge fruitful discussions with R. Bruinsma, F. J\"ulicher, F.
K\'ep\`es, J.-M. Delosme, B. Goud, P.  Chavrier, V. Norris, and J.-M.
Valleton.
\end{acknowledgments}

\appendix
\section{Many-body interactions between point-like
curvature sources} 
\label{app_calc}
Let us outline the derivation of the interaction between $N$ anisotropic
point-like sources that locally imprint a curvature on the membrane. As
explained in the text, such constraints modelize a wide class of
membrane inclusions, including transmembrane and cytosol proteins partly
inserted within the membrane. 

For small deformations $u(x,y)$ with respect to the $(x,y)$ plane, the
free energy associated with the curvature elasticity of a membrane is
given by~\cite{helfrich73}:
\begin{equation}\label{elnrj}
F_\mathrm{el}=\frac{\kappa}{2}\int\!dx\,dy\left(\nabla^2u\right)^2.
\end{equation}
Indeed, for small deformations, the Laplacian $\nabla^2u(\mathbf{r})$ is
equal to the sum of the membrane's principal curvatures at point
$\mathbf{r}=(x,y)$.  The material parameter $\kappa$ is the bending
rigidity. 

Determining the shape of the membrane in the presence of inclusions at
positions $\mathbf{r}_\alpha$ imprinting local curvatures requires
minimizing the elastic energy (\ref{elnrj}) with local constraints on
the membrane curvature tensor.  In the small deformation limit, the
elements of the latter are given by the second spatial derivatives of
the membrane shape: $u_{,xx}(\mathbf{r})$, $u_{,xy}(\mathbf{r})$ and
$u_{,yy}(\mathbf{r})$.  Introducing $3N$ Lagrange multipliers
$\Lambda_{ij}^\alpha$ to enforce the curvature constraints, the
Euler-Lagrange equation corresponding to the constrained minimization
is
\begin{eqnarray}
&&\nabla^2\nabla^2u\left(\mathbf{r}\right)=\sum_{\alpha=1}^N
\Big[
\Lambda_{xx}^\alpha\,\delta_{,xx}\left(\mathbf{r}-\mathbf{r_\alpha}\right)
\nonumber\\
&&\quad+\Lambda_{xy}^\alpha\,\delta_{,xy}\left(\mathbf{r}-\mathbf{r_\alpha}\right)
+\Lambda_{yy}^\alpha\,\delta_{,yy}\left(\mathbf{r}-\mathbf{r_\alpha}\right)
\Big],\qquad
\end{eqnarray}
where $\delta(\mathbf{r})$ is the two-dimensional Dirac's delta and a
comma indicates derivation. By linearity, the solution of this equation
is
\begin{equation}\label{sol}
u(\mathbf{r})=\sum_{\mu=1}^{3N}\Lambda_\mu \Gamma_\mu(\mathbf{r}),
\end{equation}
where the $\Lambda_\mu$'s and $\Gamma_\mu$'s are the $3N$ components of the
column matrices
\begin{equation}
\mathsf{\Lambda}=\pmatrix{\Lambda_{xx}^1\cr\Lambda_{xy}^1\cr
\Lambda_{yy}^1\cr\Lambda_{xx}^2\cr\vdots},
\qquad
\mathsf{\Gamma}(\mathbf{r})=\pmatrix{
G_{,xx}(\mathbf{r}-\mathbf{r}_1)\cr
G_{,xy}(\mathbf{r}-\mathbf{r}_1)\cr
G_{,yy}(\mathbf{r}-\mathbf{r}_1)\cr
G_{,xx}(\mathbf{r}-\mathbf{r}_2)\cr
\vdots},
\end{equation}
and
\begin{equation}
G(\mathbf{r})=\frac{1}{16\pi}r^2\ln r^2
\end{equation}
is the Green function of the operator $\nabla^2\nabla^2$, satisfying the
equation $\nabla^2\nabla^2G(\mathbf{r})=\delta(\mathbf{r})$.

We introduce a column matrix $\mathsf{K}$ containing the values of the $3N$
constraints
\begin{equation}
\mathsf{K}=\pmatrix{u_{,xx}(\mathbf{r}_1)\cr
u_{,xy}(\mathbf{r}_1)\cr
u_{,yy}(\mathbf{r}_1)\cr
u_{,xx}(\mathbf{r}_2)\cr
\vdots}.
\end{equation}
With $u(\mathbf{r})$ given by Eq.~(\ref{sol}), enforcing the constraints
yields the following equation for the Lagrange multipliers:
\begin{equation}\label{lagranges}
\sum_{\nu=1}^{3N}M_{\mu\nu}\Lambda_\nu=K_\mu,
\end{equation}
\smallskip
where the $3N\times3N$ matrix~$\sf M$ is given by
\begin{equation}\label{matrixM}
{\sf M}=\pmatrix{
{\sf m}_{11}&{\sf m}_{12}&\ldots&{\sf m}_{1N}\cr
{\sf m}_{21}&{\sf m}_{22}&&\vdots\cr
\vdots&&\ddots&\vdots\cr
{\sf m}_{N1}&\ldots&\ldots&{\sf m}_{NN}
},
\end{equation}
in which the ${\sf m}_{\alpha\beta}$'s are $N^2$ matrices of size
$3\times3$ defined by 
\begin{equation}
\label{Gxxxx}
{\sf m}_{\alpha\beta}\!=\!\pmatrix{
\!G_{,xxxx}\left({\bf r}_{\beta\alpha}\right)\!
&
\!G_{,xxxy}\left({\bf r}_{\beta\alpha}\right)\!
&
\!G_{,xxyy}\left({\bf r}_{\beta\alpha}\right)\!
\cr
\!G_{,xxxy}\left({\bf r}_{\beta\alpha}\right)\!
&
\!G_{,xxyy}\left({\bf r}_{\beta\alpha}\right)\!
&
\!G_{,xyyy}\left({\bf r}_{\beta\alpha}\right)\!
\cr
\!G_{,xxyy}\left({\bf r}_{\beta\alpha}\right)\!
&
\!G_{,xyyy}\left({\bf r}_{\beta\alpha}\right)\!
&
\!G_{,yyyy}\left({\bf r}_{\beta\alpha}\right)\!
},
\end{equation}
where $\mathbf{r}_{\beta\alpha}=\mathbf{r}_\alpha-\mathbf{r}_\beta$.
Setting
\begin{equation}
{\bf r}_\alpha-{\bf
r}_\beta=r_{\beta\alpha}\left[\cos\theta_{\alpha\beta}\,{\bf\hat
x}+\sin\theta_{\alpha\beta}\,{\bf\hat y}\right],
\end{equation}
yields explicitely
\begin{widetext}
\begin{equation}
\label{matricesmij}
{\sf m}_{\alpha\beta}=\frac{1}{4\pi r_{\beta\alpha}^2}\pmatrix{
\cos\left(4\,\theta_{\alpha\beta}\right)-2\cos\left(2\,\theta_{\alpha\beta}\right)
&
\sin\left(2\,\theta_{\alpha\beta}\right)\left[2\cos\left(2\,\theta_{\alpha\beta}\right)-1\right]
&
-\cos\left(4\,\theta_{\alpha\beta}\right)
\cr
\sin\left(2\,\theta_{\alpha\beta}\right)\left[2\cos\left(2\,\theta_{\alpha\beta}\right)-1\right]
&
-\cos\left(4\,\theta_{\alpha\beta}\right)
&
-\sin\left(4\,\theta_{\alpha\beta}\right)-\sin\left(2\,\theta_{\alpha\beta}\right)
\cr
-\cos\left(4\,\theta_{\alpha\beta}\right)
&
-\sin\left(4\,\theta_{\alpha\beta}\right)-\sin\left(2\,\theta_{\alpha\beta}\right)
&
\cos\left(4\,\theta_{\alpha\beta}\right)+2\cos\left(2\,\theta_{\alpha\beta}\right)
}.
\end{equation}
\end{widetext}

Integrating Eq.~(\ref{elnrj}) by parts and taking into account the
constraints yields the elastic energy
\begin{equation}
\label{interaction}
F_\mathrm{el}=\frac{1}{2}\kappa\,\mathsf{K}^t \mathsf{M}^{-1}\mathsf{K},
\end{equation}
where $\mathsf{K}^t$ is the transpose of~$\mathsf{K}$. From Eqs.\
(\ref{sol}) and~(\ref{lagranges}), the equilibrium shape of the membrane
is given by
\begin{equation}\label{ueq}
u(\mathbf{r})=\mathsf{K}^t \mathsf{M}^{-1}\mathsf{\Gamma}(\mathbf{r}).
\end{equation}

When $\alpha=\beta$, ${\sf m}_{\alpha\beta}$ as given by
Eq.~(\ref{matricesmij}) diverges: indeed Eq.~(\ref{elnrj}) correctly
describes the membrane elastic energy only for distances $r\gtrsim r_0$,
where $r_0$ is of the order of the membrane thickness. It is therefore
necessary to introduce a high wavevector cutoff $r_0^{-1}$ in the
theory.  From the definition of the Green function $G({\bf r})$, we
deduce, in Fourier space
\begin{equation}
G_{,xxxx}({\bf r})=\int\!\frac{d^2q}{(2\pi)^2}
\frac{q_x^4\,e^{i{\bf q}\cdot{\bf r}}}{q^4}.
\end{equation}
Hence, introducing the cutoff, we obtain
\begin{equation}
G_{,xxxx}({\bf 0})=\int_0^{r_0^{-1}}\frac{q\,dq}{(2\pi)^2}
\int_0^{2\pi}d\theta\,\cos^4\theta=\frac{3}{32\pi r_0^2},
\end{equation}
and similarly for the other elements of the matrix~(\ref{Gxxxx}).
With the above prescription, we obtain
\begin{equation}\label{matricesmii}
{\sf m}_{\alpha\alpha}=\frac{1}{32\pi r_0^2}\pmatrix{
3&0&1\cr0&1&0\cr1&0&3}.
\end{equation}

As an illustration, let us consider the case of two identical isotropic
inclusions, each prescribing the curvature~$c$. Then
\begin{equation}
\mathsf{K}^t=\left(c,0,c,c,0,c\right),
\end{equation}
and, from Eqs.\ (\ref{interaction}), (\ref{matrixM}),
(\ref{matricesmij}), and~(\ref{matricesmii}), with $r_{12}=R$, the
interaction energy is
\begin{equation}
\label{2iso}
F_\mathrm{el}=\frac{512\,\pi\kappa\left(r_0 c\right)^2}%
{\left(\displaystyle\frac{R}{r_0}\right)^4
+8\left(\displaystyle\frac{R}{r_0}\right)^2-32},
\end{equation}

\smallskip\noindent
in which we have discarded a constant term. Setting $r_0=a/2$, we indeed
obtain the leading asymptotic interaction~(\ref{Fclcl}). This special
choice for $r_0$ allows to match the result of Goulian et al.\ (1993),
which was obtained from multipolar expansions.  It should be noted that
the interaction given by Eq.~(\ref{2iso}) is \textit{exact} within the
present formalism (for $r$ larger than~$\simeq\!a$), whereas
multipolar expansions can only give in analytical form the leading
asymptotic orders.

When many inclusions are present, the matrix $\sf M$ and its inverse,
which determines the interaction energy through Eq.~(\ref{interaction}),
can be easily calculated numerically once the positions of the
inclusions are defined.

\section{Monte Carlo simulations}
\label{app_mcarl}

The Monte Carlo simulation that we perform employs the standard
Metropolis algorithm~\cite{metropolis53}. For given positions and
orientations of the particles representing the dynamins, the energy is
numerically calculated from Eq.~(\ref{interaction}). At each Monte Carlo
step, we perform a Metropolis move consisting in either a translation or
a rotation of one arbitrarily chosen dynamin particle. The amplitude of
the moves is adjusted in order to keep an average acceptance rate of
$50\%$. We confine the dynamins inside a circular box (of radius
$80\,a$) centered around the clathrin lattice, which is kept fixed. To
take into account the hard-core repulsion (see Sec.~\ref{collective}),
we simply reject any move bringing two particles closer than the minimum
approach distance $d$ (here $2a$). Note that in this simulation the
membrane is not discretized: the interaction energy that we use fully
takes into account the elasticity of the membrane.


\end{document}